# High Contrast Cleavage Detection

Michael Dubrovsky,[1,2*] Morgan Blevins[2], Svetlana V. Boriskina[2*], Diedrik Vermeulen[1]

[1]*SiPhox Inc., 325 Vassar Street, Cambridge, MA 02139, USA*
[2]*Massachusetts Institute of Technology, 77 Massachusetts Avenue, Cambridge, MA 02139, USA*
*Corresponding authors: Mike@siphox.com , sborisk@mit.edu

**Photonic biosensors that use optical resonances to amplify signals from refractive index changes offer high-sensitivity, real-time readout, and scalable, low-cost fabrication. However, when used with classic affinity assays they struggle with noise from non-specific binding and are limited by the low refractive index and small size of target biological molecules. In this letter, we introduce the High Contrast Cleavage Detection (HCCD) mechanism, which makes use of dramatic optical signal amplification caused by the cleavage of large numbers of high-contrast nanoparticle reporters instead of the adsorption of labeled or unlabeled low-index biological molecules. We evaluate the advantages of the HCCD detection mechanism over conventional target-capture detection techniques when using the same label and the same photonic biosensor platform and illustrate numerically the possibility for attomolar sensitivity for HCCD using an example of a silicon ring resonator as an optical transducer decorated with silicon nanoparticles as high-contrast reporters. In the practical realization of this detection scheme, detection specificity and signal amplification can be achieved via collateral nucleic acid cleavage caused by enzymes such as CRISPR Cas12a and Cas13 after binding to a target DNA/RNA sequence in solution.**

The need for a step-change in sensing technology has been highlighted by the ongoing SARS-CoV-2 pandemic. Standard approaches to resonant-surface optical biosensing, highlighted in Fig. 1(a-c), rely on the affinity between the target analyte and biomolecules on the surface of the sensor [1–5]. The sensitivity of any direct unlabeled readout is limited by the small difference in optical properties between water (with refractive index of 1.33 in the visible range) and biomolecules (typically having a refractive index of ~1.45), while the selectivity is limited by the comparable optical response induced by the non-specific adsorption of non-target biomolecules to the surface [6]. Although a variety of anti-fouling coatings are used to limit non-specific binding and reduce noise in complex media [7], and high-index reporters such as beads, metal nanoparticles or fluorophores can be added to the target of interest to increase the optical signal [8] (Fig. 1(c)), these solutions add cost and complexity, and are imperfect. Antifouling coatings diminish nonspecific binding to the surface of the sensor, but they do not address the off-target binding between receptors such as antibodies or DNA strands attached to the surface of the sensor and biomolecules in the sample. This problem is magnified with the addition of reporters to the system, which can also bind off-target directly to the capture molecules on the sensor surface, requiring careful choice (often via trial and error) of the reporter and the capture molecule [9]. These issues have limited the commercial use of resonant optical biosensors in low-analyte concentration pathogen detection in favor of biological machinery, such as the enzymes used for Polymerase Chain Reaction assays, which routinely operate near single target molecule per microliter concentrations and with very high selectivity. To take advantage of the optical signal amplification in resonant photonic sensor platforms, as well as the specificity and biological amplification offered by enzymatic approaches, we propose High Contrast Cleavage Detection, as illustrated in Fig 1(e), which relies on the cleavage of multiple high index reporters from the surface of a resonant biosensor by a cleavage complex (typically an enzyme) after it is activated by the target analyte in solution. The detection of enzymatic cleavage activity (already demonstrated in SERS platforms [10,11]) is enhanced with high contrast reporters and paired with enzymes that have analyte-dependent collateral reporter cleavage activity. HCCD has three major advantages: (1) Specificity is derived from the cleavage complex, which can be chosen to be a highly specific enzyme such as one of the CRISPR-Cas12 and Cas13 systems used in viral diagnostics [12,13] and shown to cleave DNA-tethered nanoparticles from a solid substrate [14], a DNA-zyme [15] or restriction enzymes generated in toehold switch assays [16], (2) Universal reporters can be developed to work with several optical biosensor platforms and different pathogens (both existing and yet unknown), reducing the time and cost of initial sensor development, (3) Each activated cleavage complex can cleave multiple reporters, providing multiplicative optical signal amplification and reducing the limit of detection (LOD). In the case of nucleic acid detection, previous work has shown that CRISPR Cas12a and Cas13 biological machinery can be harnessed to detect specific viral RNA or DNA strands [12,13], including recent efforts that have successfully shown SARS-CoV-2 detection [17,18] and are in late stages of commercialization. These techniques typically rely on measuring changes in the fluorescent response of reporters (achieved via DNA/RNA oligo-probe cleavage to separate a quencher and a fluorophore as seen in Fig. 1(d)), and thus require nucleic acid pre-amplification (e.g. PCR, LAMP) to reach attomolar sensitivity [19]. In principle, CRISPR-enabled single molecule detection is feasible, as a single CRISPR complex, activated by the hybridization of viral RNA/DNA to CRISPR RNA/DNA, produces thousands of non-specific collateral cleavages [20]. However, fluorescent readout sets the lower limit for sensitivity in the fM range [12], and the nucleic acid pre-amplification process requires 1-3 hours of sample preparation and incubation, preventing real-time rapid screening. Seen in Fig. 2 below, HCCD provides a new method for developing CRISPR assays enhanced by the photonic signal amplification mechanism.

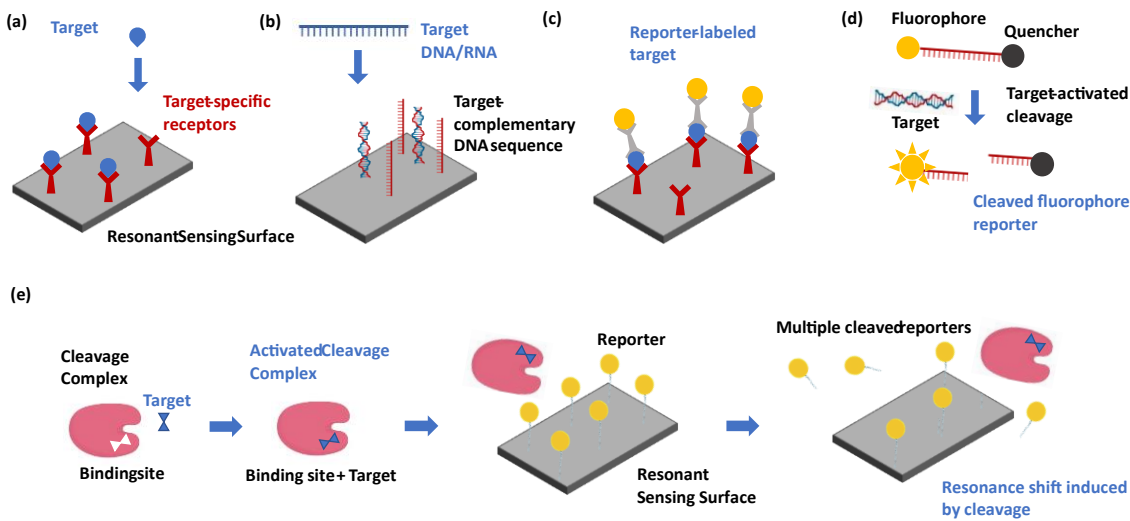

**Fig. 1.** (a-b) Label-free receptor-affinity capture based detection mechanism. (c) Capture-based sensor optical signal can be amplified by using target-specific labels or sandwich-type assays. (d) RT-PCR and other techniques based on biological target amplification make use of target-activated fluorescent reporter cleavage for optical detection. (e) HCCD makes use of the amplification mechanism based on multiple reporters collaterally cleaved by each cleavage complex activated by the target. Reporters and receptors are not target-specific and can be used to detect different and new pathogens.

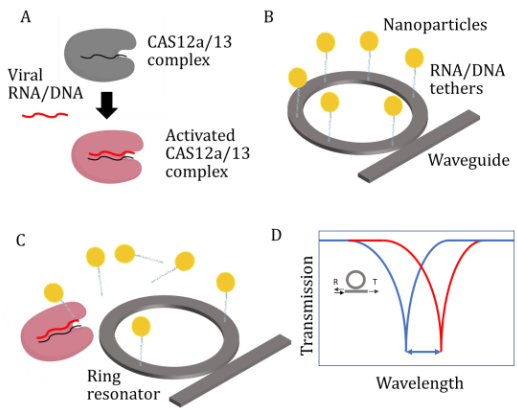

**Fig. 2** CRISPR-HCCD in a resonant mirroring sensor platform begins with (a) the activation of a CRISPR complex by a target matching the guide-RNA carried by the CRISPR complex, the CRISPR complex then diffuses to the surface of an optical micro-ring resonator decorated with ss-DNA-Nanoparticle reporters (b). The activated CRISPR complex cleaves multiple reporters non-specifically (c), resulting in a detectable resonance shift in the transmission spectrum (d) of the ring.

We will illustrate the advantages of the proposed HCCD mechanism with an example of an integrated microresonator [21] (MRR) photonic biosensor (Fig. 2), which offers a highly sensitive readout via monitoring spectral shifts and intensity changes of narrow spectral resonances associated with excitation of resonator modes [22–25]. Since previous work showed that attachment of multiple metal nanoparticles to the microring surface can severely degrade its Q-factor [26,27], we have chosen Si nanoparticle reporters in the following calculations. First, we used the Finite Difference Eigenmode solver (Lumerical MODE package) to calculate the spectral shift of a transverse-electric (TE) WG mode of a 20 micron diameter microring in the 1.55-micron band, which is caused by the random cleavage of tethered reporters (Fig. 3). Figure 3 (a) illustrates the difference in the WG mode electric field profile in the microring cross-section before and after five 40nm diameter Silicon nanoparticles (SiNP) are cleaved from its surface. The resulting shift of the resonance frequency of the microring as a function of the number of reporters cleaved was analytically determined based on the changes in the effective index of the simulated modes (Fig. 3 (b)). This simulation was repeated by varying the number and position of SiNP reporters while keeping them between 10nm and 20nm away from the surface of the waveguide to simulate realistic conditions, which provided an estimate of the standard deviation of the resonant shift (shaded blue area). Additional simulations were conducted, showing in Fig. 3(c) the dependence and standard deviation of the resonance shift on the size of a single SiNP attached to the waveguide surface in a random position. Notably, the standard deviation increases with the particle size, meaning that replacing large reporter particles with multiple smaller ones may give data that would be easier to use in a quantitative test. 40 nm diameter SiNPs offer a compromise between the resonant shift amplitude and randomness-induced standard deviation.

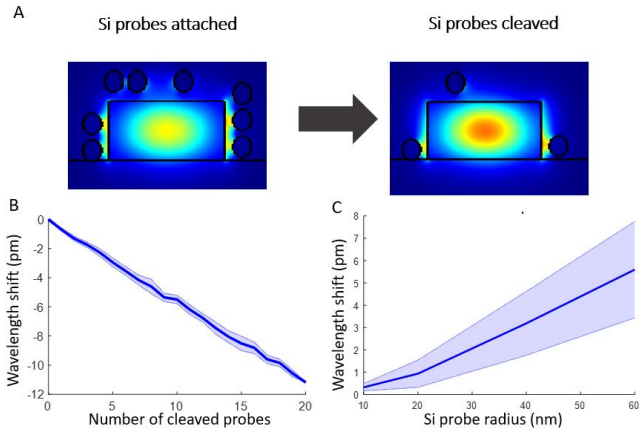

**Fig. 3** (a) Electric field intensity profiles for the fundamental TE mode in a 500 nm wide and 220 nm thick strip silicon on insulator (SOI) waveguide in water before and after cleavage of five 40nm-diameter SiNP reporters. (b) WG mode wavelength blue shift as a function of number of 40nm-diameter SiNP reporters removed. (c) WG mode wavelength redshift caused by attachment of a single SiNP as a function of NP size. In both (b) and (c) a standard deviation is shown based on variation of the distance of the particles to the waveguide surface from 10 to 20 nm and the variation of the position of the particles around the waveguide.

To further investigate the effect of the cleavage of multiple SiNP reporters from a microring, we performed three-dimensional finite difference time domain calculations of transmission spectra of a 40 micron diameter Si micro-ring resonator side-coupled to a bus Si waveguide via an 120nm gap (FDTD Lumerical package). The microring is initially decorated by three thousand 40nm-diameter SiNP reporters and is excited by the fundamental TE guided mode of the bus waveguide. The signal transmitted through the waveguide exhibits narrow spectral dips at the wavelengths corresponding to WG mode resonances in the ring, which blue shift progressively as more particles are removed from the microring surface (Fig. 4b), as would happen during cleavage under experimental conditions. The position of the particles was again randomly varied while keeping the distance to the surface of the waveguide at 15nm and the standard deviation of the cleavage-induced resonance shift can be seen to decrease as a greater number of particles are cleaved. We note that the average wavelength shift for a single 40nm diameter SiNP reporter removal (-0.56 pm) predicted in our MODE simulations agrees qualitatively with the shift predicted by the 3D FDTD simulations (-0.30pm). Importantly, this shift is approximately the minimum shift that can be detected experimentally by using standard microring sensor architectures [28]. Moreover, this blue-shift is opposite to the redshift caused by non-specific binding, meaning that attomolar sensitivity in complex samples may be achievable. In order to optimize the design of an MRR-HCCD sensor, it is useful to examine the factors that govern its LOD in contrast with those that govern the LOD of a classic MRR affinity sensor.

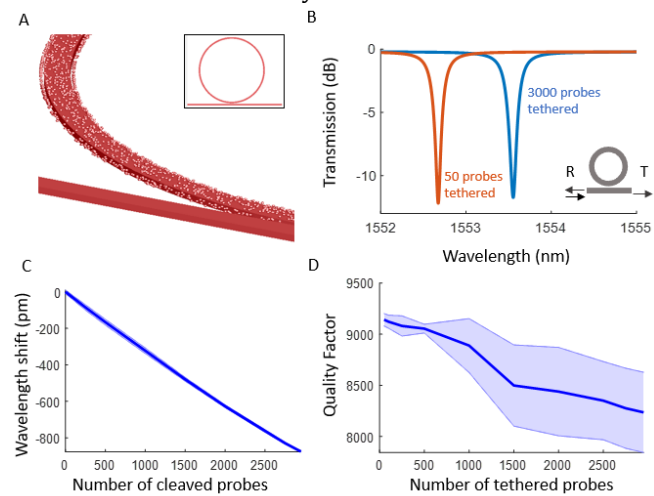

**Fig. 4** (a) Schematic of a 20 micron diameter micro-ring resonator side-coupled to a bus Si waveguide with the same parameters as in Fig. 2, initially decorated by 3000 40 nm diameter SiNP reporters arranged randomly around the ring circumference 15 nm away from its surface. (b) The resonator spectrum shifts as the number of probes is reduced to 50; the blue-shift of one of the WG mode resonances after cleavage is shown. (c) The wavelength shift for the spectral dip in panel b as a function of the number of cleaved Au reporters, showing the standard deviation based on variation of the distance of the reporters to the waveguide surface from 10 to 20 nm and the variation of the random positions of the reporters around the ring. (d) Q factor as a function of number of particles attached to the surface of the resonator

The bulk refractive index LOD of resonant optical biosensors is given by the equation $LOD = \frac{R}{S}$ where $R$ is the sensor resolution (depending on spectral noise, temperature noise, and amplitude noise as well as the sensor optical mode quality factor) and $S$ is sensitivity to bulk refractive index changes [29,30]. In the case of discrete reporter binding or reporter cleavage events, we must also consider the percentage of the surface that is available for binding or cleavage and the refractive index and size of the binding or cleaved particles. A relative comparison of a sensor operated in HCCD mode (Fig. 1(e)) vs. affinity mode (Fig. 1 (a-c)) reveals that an HCCD sensor, assuming the same reporter particle in both cases, has the advantage of multiple reporter events per detected target, but suffers from Q-factor degradation (leading to amplitude noise) due to the attachment of high-contrast reporters to the surface which introduce loss in the waveguide (whereas the affinity sensor only requires a surface of low-index receptor biomolecules). This tradeoff suggests that there is an optimal surface coverage for HCCD sensors depending on the spectral resolution of the source and other considerations specific to a given implementation, and we see from simulations (Fig. 4 (d)) that the Q factor decays linearly from ~9100 to ~8250 as 3000 particles are attached covering ~8% of the ring's surface area. Given the same sensor and the same reporter particle, the advantages of the HCCD approach come from the multiplication factor associated with the detection enzyme's activity as well as the improved diffusivity of the detection enzyme (typically a protein around 3nm in diameter) as compared to reporter particles which range up to ~1 micron in diameter.

| Sensor Type | Affinity | HCCD |
|---|---|---|
| Sensitivity to a single reporter | Equivalent | Equivalent |
| Number of reporters bound/cleaved per target | Single | Multiple |
| Surface coverage with receptor biomolecules (affinity) or reporter probes (HCCD) | Maximum allowed by steric hinderance | Tradeoff with Q-factor degradation, to be optimized for each system |
| Diffusion Effects | Depends on the diffusion of large reporter | Depends on the diffusion of small cleavage complex/enzyme |

Due to the significant shift that can be achieved even from the cleavage of a single reporter, the practical limits of detection and time response in such a system may be determined by the diffusion rate of activated CRISPR complexes to the probe cleavage sites [31,32]. Silicon photonic microring sensors have been demonstrated to be very scalable via production in standard silicon photonic processes [28,33], however, the HCCD approach is completely general, and can be combined with other optical transducers (e.g. surface plasmon resonance sensors, bioluminescence imaging sensors or integrated Mach-Zehnder interferometers), and offers a pathway to highly sensitive real-time detection of biological molecules, viruses and other microorganisms. Any biosensing architectures based on analyte-induced enzymatic cleavage such as DNA-zyme biosensors [34] for sensing metal ions and Toehold Switch RNA sensors [35] are promising candidates for enhancement via HCCD readout and we hope the concept of communication between on-chip devices and biological systems via HCCD can be further generalized as opto-biological nano-systems become more prevalent.

**Acknowledgments**. We thank Ke Du at RIT, John Connor at BU, Sharon Weiss at Vanderbilt University and Peter Nyugen at the Wyss Institute for helpful discussions as well as Mustafa Hammood at UBC for assistance with the FDTD simulation setup.
**Disclosures**. SiPhox has patents pending on this technology.